\documentclass[9pt,twocolumn]{article} 
\usepackage{simpleConference}
\usepackage{times}
\usepackage{graphicx}
\usepackage{amssymb}
\usepackage{url,hyperref}

\begin{document}

\title{Controlling group and phase velocities in bidirectional mode-locked fiber lasers}

\author{Hanieh Afkhamiardakani, Jean-Claude Diels \\
\\
\small Center for High Technology Materials, University of New Mexico, Albuquerque, NM 87106, USA \\
\small \today
\\
\em \small corresponding author: jcdiels@unm.edu  \\
}

\maketitle
\thispagestyle{empty}

\begin{abstract}
\textbf { A bidirectional mode-locked fiber laser producing two correlated frequency combs of the
same repetition rate is demonstrated.
The intensity dependence of pulse and phase velocities  are measured simultaneously.
As expected, the phase delay is determined by the linear and nonlinear indices of  refraction.
The nonlinear Kerr effect contribution to the pulse velocity is dwarfed
by a contribution from saturable gain dynamics. }
\end{abstract}

\section{Introduction}
Fiber technology is traditionally used for sensing applications,
where a phase shift is measured either in a resonator or
through evanescent wave coupling in a tapered fiber.  For instance,  Sagnac interference
in an optical system consisting of a fiber loop with counter-propagating
light beams can be used for rotation, temperature, tension,
magnetic field sensing~\cite{Shao16,Naeem15,Lv15} and
medical applications~\cite{Kumagai14}. Michelson and Fabry-Perot
interferometers are also used  for temperature sensing~\cite{Li14,Yu18}.
A simple structure based on non-adiabatic tapered optical fiber
 utilizes the evanescent field of an optical microfiber
 to enhance the light-biomaterial interaction for biosensing
 applications~\cite{WangPengfei15, Lin12}.

A considerable increase in sensitivity can be achieved in all phase sensing applications
by inserting the element to be measured inside an active laser cavity.
In the technique of Intracavity Phase Interferometry (IPI), a mode-locked laser with two intracavity counter-propagating pulses produces two correlated frequency combs without need of stabilization.
Interfering these two combs produces a beat note
 frequency
proportional to the phase shift per cavity round-trip~\cite{Arissian14b}.
 The bandwidth
of the beat note can be as small as 0.2 Hz~\cite{Velten10}, even though each comb has a bandwidth larger than 1 MHz, again indicating the correlation between combs.
The beat note produced by the interference of the two combs can be expressed as~\cite{Arissian14b}:
\begin{equation}
\Delta \nu = \nu \frac{\Delta \varphi}{P k_{av}} =  \nu \frac{\Delta P}{P},
\label{deltanu}
\end{equation}
where $\nu$ is the optical frequency, $\Delta \varphi$ is the induced
phase shift (to be measured) per cavity round-trip, $k_{av}$ is the averaged $k$ vector over the cavity, $\Delta P$ is the difference in optical path length
that would correspond to the phase shift, and $P$ is the optical perimeter of the laser.

The IPI technique has successfully made measurements of
nonlinear index, gas flow, electro-optic coefficients, and rotation~\cite{Velten10,Dennis91b,Bohn97b,Lai92c}.
These demonstrations used free-space component lasers which do not lend themselves to field applications.
Mode-locked fiber lasers are the most promising media to
implement IPI due to their ability to produce ultrashort pulses in a compact design.

There is abundant literature on {\em unidirectional} mode-locked ring fiber lasers~\cite{Duling91,Nishizawa08,Set04,Song07,Bao09,Sotor15}.
Stable frequency combs have also been realized with linear lasers~\cite{zhang09b,guo14}. However,
achieving colliding pulse mode-locking~\cite{Bradley76,Fork81} in bidirectional fiber lasers to realize correlated frequency combs is
surprisingly challenging.
 The difficulties stem from the high gain and loss and large nonlinearities of fiber optics.
Pulses circulating in opposite directions in a bidirectional laser
traverse the optical components in a different order creating an asymmetry in the cavity.
This asymmetry, at the worst, leads to the tendency of the two circulating pulses to unlock from crossing in the saturable absorber. This results in having different
repetition rates (pulse velocities) and wavelengths (frequencies); a mode of operation
totally inadequate for IPI~\cite{Zeng13,Yao14,Li18}.
Remarkably successful implementations of IPI in passively mode-locked bidirectional
fiber lasers, as laser gyros, have been demonstrated by Kieu et al~\cite{Kieu08}, and more recently by Krylov et al~\cite{Krylov17}.
However, a large bias beat note is measured in fiber laser gyros as a result of the asymmetry
in phase velocities.

A nearly symmetric design of bidirectional mode-locked fiber laser is presented that has the ability to control the amount of bias beat note (phase velocity) by tuning the pump powers.
The phase velocity is determined by the linear and nonlinear refractive indices of the fiber.
% In addition,
The measurements confirm the conclusion of prior results obtained with free-space component lasers~\cite{Masuda16},
that the pulse velocity in the cavity is dominated by gain dynamics and it is not simply equal to $d\Omega/dk$,
{ generally referred to as group velocity, where $k$ is the wave vector
and $\Omega$ is the optical frequency.}

\section{Experimental setup}
\begin{figure}[t]
\centering
\includegraphics[width=\linewidth]{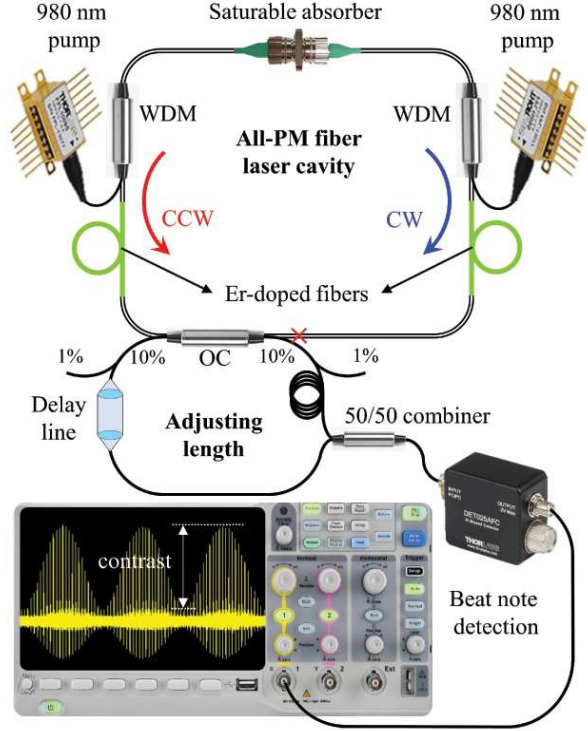}
\caption{\small Experimental setup of an all-polarization maintaining bidirectional fiber laser with minimized asymmetries using two
sections of Er-doped fibers; WDM: Wavelength Division Multiplexer, OC: Output Coupler. Saturable absorber is a thin layer of carbon nanotubes between two fiber
 ferrules which establishes the crossing point of pulses.
The other crossing point is located at the opposite side of the ring marked by a cross ($\times$).
Placed at that location, OC would cause back scattering that injection locks the two counter-circulating pulses.  To avoid the large resulting dead band,
OC is moved away from $\times$.  An adjustable delay line images the
crossing point $\times$ onto the beat note detector. }
\label{setup}
\end{figure}

An all-polarization maintaining bidirectional fiber laser is constructed and shown in Fig.~\ref{setup}. Passive Mode-locking is achieved by
sandwiching carbon nanotubes (CNTs) between two FC/APC fiber connectors~\cite{Nicholson07b} creating two correlated counter-propagating frequency combs.
 The laser cavity elements are arranged as symmetrically as possible with respect to the
saturable absorber (SA), to ensure that counter-circulating pulses of near equal intensities are generated.
The SA establishes the crossing point of
the counter-circulating pulses. A tapered fiber covered with CNTs~\cite{Afkhamiardakani16} fails to stabilize the crossing point
possibly because the taper is longer than the pulse length,
resulting in the mutual saturation being applied to only a small fraction of the total absorption.
An important contribution to minimize the asymmetries in the cavity is  to use two portions of Er-doped fibers pumped
through two wavelength division multiplexers (WDM) [Fig.~\ref{setup}].
Using WDMs in reflection helps to protect the saturable absorber from overheating by filtering out the extra power from the pump lasers,
 which also makes the mode-locking more stable~\cite{Afkhamiardakani18A}.
A 2 by 2 output coupler extracts 10\% of the light from either direction.
The output pulse trains from clockwise (CW) and counter-clockwise (CCW) directions
combine through a 50/50 combiner to measure the beat note.
As can be seen in the setup, the output coupler is not placed in exactly equal distances from SA, creating some asymmetries which can be compensated by tuning the pump powers of the two gain sections.
The near symmetric operation reduces the bias beat note due to creating less differences between phase velocities of the circulating pulses. A minimum bias frequency of 164 kHz could be achieved.

 An adjustable delay line ensures temporal overlap of the
counter-propagating pulses on the detector. The delay line is made of two collimating lenses with adjustable collimated optical length.
 1\% of the the output coupler (OC)
in either direction is splitted out and monitored for extra measurements such as output power, optical spectrum and pulse train.

\section{Experimental and theoretical results}
\subsection{Bias beat note}
\begin{figure}[t]
\centering
\includegraphics[width=\linewidth]{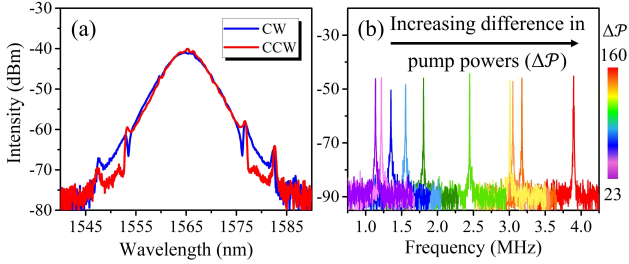}
\caption{\small Experimental results; (a) Spectra  of the pulse train for each
direction of circulation. (b) Radio frequency spectra recorded for different pump
power differences ranging from 23 mW to 160 mW.}
\label{spectrum}
\end{figure}

Colliding pulse mode-locking in a bidirectional ring fiber laser creates two counter-propagating pulses which meet at the same locations established by the saturable absorber at each round trip.
The soliton spectra in CW and CCW directions are shown in Fig. \ref{spectrum} (a) with the same central wavelength of 1565 nm for both directions.
Each pulse of intensity $I$ circulating in the cavity of perimeter $P$ accumulates a large amount of nonlinear phase $\varphi_{nl}$
along the core of the fiber given by:
\begin{equation}
\varphi_{nl} = \frac{2\pi}{\lambda} n_2 I P,
\label{phase}
\end{equation}
where $n_2$ is the nonlinear refractive index of the fiber core,  and $\lambda$ is the operating wavelength.
The difference between the accumulated phase in either direction {($\Delta \varphi$)}
 per cavity round-trip results in a beat note measured by interfering the frequency combs.
The frequency $\Delta \nu$ of that beat note is found by combining Eqs.~(\ref{deltanu}) and
(\ref{phase}):
\begin{equation}
\Delta\nu=\nu\frac{n_{2}}{n_{av}}\Delta I,
\label{btn}
\end{equation}
where $n_{av}$ is the average linear phase index.
As can be seen, the beat note frequency is directly proportional to the difference of pulse intensities of $\Delta I$. %and perimeter of $P$ =5.476 m.
Thanks to the symmetric design of the laser cavity (Fig.~\ref{setup}), each pulse sees the components in the same order starting from
the crossing point in the CNT saturable absorber, resulting in a very small difference in accumulated nonlinear phase.
One expects to measure a very small bias beat note for a perfectly symmetric cavity.
Unfortunately, making a perfect symmetric cavity is not practical as it needs the 2 $\times$ 2
output coupler (OC) placed at the crossing point (red $\times$ in Fig.~\ref{setup}) opposite to the saturable absorber.
This introduces a large coupling between the counter-circulating pulses, resulting
in a large dead band (mutual injection locking).
Therefore, the symmetry had to be broken by locating the 2 $\times$ 2 output coupler
away from the pulse crossing point ($\times$),
as  shown in Fig.~\ref{setup}.

The difference in intensity $\Delta I$ between counter-circulating pulses can be modified by tuning
the power of the two pump lasers in Fig.~\ref{setup}, resulting in a change of beat note indicated by Eq.~(\ref{btn}).
Figure~~\ref{spectrum}(b) shows the position of the beat note spectrum on the a radio frequency (RF) spectrum analyzer, as the difference between the pump powers is increased.
In this figure, the beat note frequency taken from RF spectrum analyzer shows an increase
from 1.1 MHz to 3.9 MHz by increasing the asymmetry in the cavity through changes in pump powers.

\begin{figure}[h!]
\centering
\includegraphics[width=\linewidth]{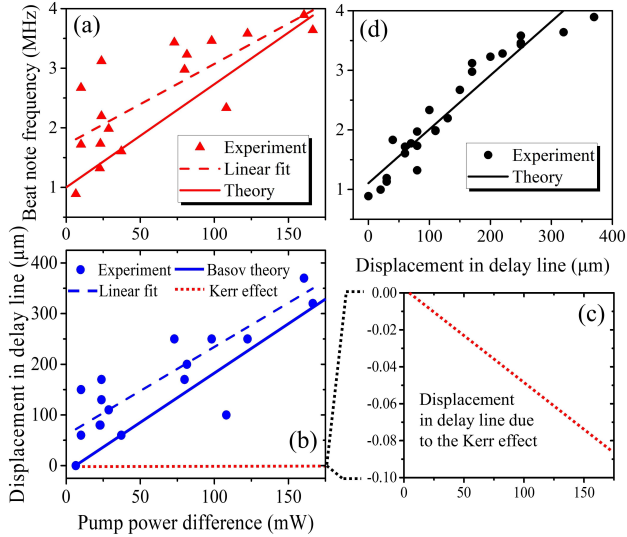}
\caption{\small Experimental and theoretical results; (a) pump power dependence of the beat note frequency in experiment (red triangles) and theory (solid red line),
(b) Displacement in delay line to achieve the maximum contrast when changing the pump powers in experiment (blue circles)
and based on the theory of Basov~\cite{Basov66} (solid blue line) which is compared with the amount of displacement based on the Kerr effect (dotted red line),
(c) magnified scale of the Kerr effect in graph (b) to show the actual behavior of displacement based on the Kerr effect,
(d) a tight correlation between beat note frequency and displacement in delay line is illustrated experimentally (black circles) and theoretically (black line).
\em {Pump power difference = (Pump power in the CW direction)-( Pump power in the CCW direction).}}
\label{results}
\end{figure}

Figure~\ref{results} illustrates the power dependence of the beat note (representative of the phase velocity)
and the delay adjustment (representative of the {pulse} velocity) to achieve optimum beat note visibility.
 The RF bias beat note is plotted in Fig.~\ref{results}(a) as a function of the
difference in pump powers ($\Delta \mathcal{P}$) applied to the two erbium doped fibers.
The difference in pump powers is calculated as $\Delta \mathcal{P}=\mathcal{P}_{cw}-\mathcal{P}_{ccw}$ where $\mathcal{P}_{cw}$ and $\mathcal{P}_{ccw}$ are the pump powers in CW and CCW directions, respectively and $\mathcal{P}_{cw}>\mathcal{P}_{ccw}$.

{The change of the bias beat note as a function of intensity can  be easily calculated from Eq.~(\ref{btn}).
A simple one to one correspondence between pump power and peak intensity can be established, under the
assumption that the laser generates only a single pulse/cavity round-trip, and that the pump power dependence of
the pulse duration can be neglected.
The
measured pulse width of 0.7 ps, repetition rate $c/(n_{av}P)    $ of 37.255 MHz, and fiber core area of $8.659 \times 10^{-7}$ cm$^2$ (corresponding to mode field diameter of $10.5 \mu m$)  have been used for both pulses,
as well as a linear dependence of the intracavity power on the pump power to calculate the intensities in either direction.
The nonlinear refractive index of $n_2 = 3 \times 10^{-16}$ cm$^2$/W at a wavelength of $\lambda$ = 1565 nm is used to calculate the beat note frequency.}
The calculated power dependence of the bias beat note under these assumptions is plotted as a { red line in Fig.~\ref{results}(a)}.
A linear fit of the experimental data is shown as dashed line in this figure. The large scatter of the experimental data is a clear indication
that the {assumptions are not accurate}.
Indeed, satellite pulses are observed for some values of pump powers {proving} that the energy of the most intense
pulse is less than the ratio of the average power to the repetition rate.

\subsection{Delay line adjustment}
If the pulses circulating in opposite direction experience a different index of refraction,
one would also expect a change in  position of the second crossing point (red $\times$ in Fig.~\ref{setup}), requiring an
adjustment of the detection delay line for maximum beat note visibility.   Blue circles in Fig.~\ref{results}(b)
show that indeed, the change in delay depends on the power difference applied
to the gain fibers.  These data were obtained by adjusting the delay line
 for maximum contrast in the beat note pattern shown in  Fig.~\ref{setup} at each pump powers setting.
 It should be noted that the plot refers to  changes in the delay line which is placed in the CW direction.
Here again, we attribute the apparently scattered pattern of the pump power dependence
to the fact that an increase in pump power does not imply a
proportional pulse intensity increase. The presence of satellite pulses results in an
erroneous estimate of the
main pulse intensity.

If the
observed changes were solely due to the Kerr effect,  the more intense pulse would be delayed by:
\begin{equation}
\Delta L=n_{2}(I_{cw}\ell_{cw}-I_{ccw}\ell_{ccw}),
\label{delay}
\end{equation}
where $I_{cw}$ and $I_{ccw}$ are the pulse intensities, each going through the
length of fiber from the saturable absorber (crossing point) to the 10\% output coupler
in clockwise ($\ell_{cw}$ = 2.856 m) and counter-clockwise
($\ell_{ccw}$ = 2.566 m) directions, respectively.
The dotted red line in Fig.~\ref{results}(b) and its magnified version in Fig.~\ref{results}(c) shows
the amount of delay between counter-circulating pulses due to the Kerr effect.
This delay is considerably smaller than the experimental observation, and of opposite sign.
Therefore, it can be concluded that the changes in index do not explain the intensity dependence of the
pulse velocity.
Instead, the much larger  delay observed between two pulses can be explained by the fact that the power dependence
of the group velocity {\em in an active laser} is dominated by gain dynamics~\cite{Masuda16,Hendrie16}.

\subsubsection{Ratio of beat note to displacement}
The surprising observation is that the ratio of the beat note to the change in delay
follows a linear dependence depicted as black circles in Fig.~\ref{results}(d), even though the
dependency of the beat note frequency and delay to pump powers are rather scattered.
The fact that the bias beat note and the change in delay are so correlated
 is an indication that both are proportional to the intensity of the
two pulses interfering on the detector.
The change in delay line is
 about 370 $\mu$m for a 3.1 MHz change in beat note.
It should be noted that the theory [black line in Fig.~\ref{results}(d), which is the ratio of the theory lines in
 Fig.~\ref{results}(a) and Fig.~\ref{results}(d)] perfectly fits the experiment.

The phase velocity intensity dependence is well explained by the Kerr effect [Eq.~(\ref{delay})].
It is shown below that propagation in a saturated gain medium explains the sign and magnitude of the intensity
dependence of the pulse velocity.

\subsubsection{Group velocity modification through gain dynamics}

 Basov {\em et al}~\cite{Basov66} showed that the velocity of a pulse in a saturable gain medium fits the expression:
\begin{equation}
\frac{v}{c} = 1 + \frac{c\tau_p}{2}(\alpha - \gamma),
\label{superluminal}
\end{equation}
where $c$ is the velocity of light in the medium, $\tau_p$ is the pulse duration, $\alpha$ is the small signal gain coefficient,
and $\gamma$ the loss coefficient (per unit length).
 We adapt the very simple model of Basov
to calculate the asymptotic pulse velocity in our fiber laser.  To this effect,
circulation of a pulse in a ring fiber laser of perimeter $P$ is replaced by pulse propagation through a distributed
amplifier of gain $\alpha = \alpha_0 \mathcal{P}/\mathcal{P}_0$ and distributed loss of $\gamma$.
$\mathcal{P}$ is the pump power, equal to $\mathcal{P}_0$ at threshold, where the gain $\alpha_0$ is calculated
from the threshold condition of $ R \exp(\alpha_0 \ell_g) = 1$; $R$ being the total survival factor per round-trip, and
$\ell_g$ the length of the erbium doped fiber.  The distributed loss $\gamma$ per unit length
is calculated from $R = \exp(-\gamma P)$.  For our fiber laser, we have $\alpha_0 = 2.5$ m$^{-1}$ and
$\gamma = 0.48$ m$^{-1}$.   Modifying the expression from Basov [Eq.~(\ref{superluminal})], the superluminal
pulse velocity in the fiber is given by:
\begin{equation}
v = c + \frac{\tau_p c^2}{2}(\alpha_0 \frac{\mathcal{P}}{\mathcal{P}_0}-\gamma).
\label{superluminal-fiber}
\end{equation}
The calculated delay versus pump power difference based on the theory of Basov is plotted as a solid blue line in Fig.~\ref{results}(b) which is
consistent with the experiment. {It should also be mentioned that the zero displacement in the delay line is defined as the position for which the beat note is measured for the minimum difference in pump powers.}

\subsection{Comparison of theory and experiment}
The theoretical curves of Figs.~\ref{results}(a) and (b) are assuming an ideal world where the pulse
intensity varies monotonically and proportionally to the pump power.
The experimental beat note  plotted as a function of the displacement (black dots) is compared to the theory (black line) in Fig.~\ref{results}(d).
The beat note is proportional to the variation of phase delay in the fiber laser, while the displacement is proportional to
the variation of pulse delay.
 The agreement between theory and experiment in Fig.~\ref{results}(d)
is quite remarkable, considering that there are no adjustable parameters, and
given the coarse assumptions (uniform intensity through the fiber and modeling of the ring laser as a uniform amplifier medium).
The
calculated plot of beat note versus delay adjustment [solid line in Fig.~\ref{results}(d)]
match exactly the
best fit line of the experimental data [dashed black line in Fig.~\ref{results}(d)].
It is often claimed that the pulse velocity is given simply by $1/(dk/d\Omega)$ where $k$ is the wave vector
and $\Omega$ the optical frequency.  The measurements presented here clearly demonstrate that this definition applies only to
transparent dielectrics.  Inside a laser cavity, the pulse velocity differs from this definition by orders of
magnitude, as has been shown for free-space component lasers~\cite{Hendrie16}.


\begin{thebibliography}{10}
\newcommand{\enquote}[1]{``#1''}

\bibitem{Shao16}
L.-Y. Shao, X.~Zhang, H.~He, Z.~Zhang, X.~Zou, B.~Luo, W.~Pan, and L.~Yan,
  \enquote{Optical fiber temperature and torsion sensor based on lyot-sagnac
  interferometer,} {{Sensors}} \textbf{16} (2016).

\bibitem{Naeem15}
K.~Naeem, B.~H. Kim, B.~Kim, and Y.~Chung, \enquote{Simultaneous
  multi-parameter measurement using sagnac loop hybrid interferometer based on
  a highly birefringent photonic crystal fiber with two asymmetric cores,}
  {{Opt. Express}} \textbf{23}, 3589--3601 (2015).

\bibitem{Lv15}
F.~Lv, C.~Han, H.~Ding, Z.~Wu, and X.~Li, \enquote{Magnetic field sensor based
  on microfiber sagnac loop interferometer and ferrofluid,}
  {{IEEE Photonics Technology Letters}} \textbf{27},
  2327--2330 (2015).

\bibitem{Kumagai14}
T.~Kumagai, Y.~Tottori, R.~Miyata, and H.~Kajioka, \enquote{Glucose sensor with
  a sagnac interference optical system,} {{Appl. Opt.}}
  \textbf{53}, 720--726 (2014).

\bibitem{Li14}
Z.~Li, Y.~Wang, C.~Liao, S.~Liu, J.~Zhou, X.~Zhong, Y.~Liu, K.~Yang, Q.~Wang,
  and G.~Yin, \enquote{Temperature-insensitive refractive index sensor based on
  in-fiber michelson interferometer,} {{Sensors and
  Actuators B: Chemical}} \textbf{199}, 31--35 (2014).

\bibitem{Yu18}
H.~Yu, Y.~Wang, J.~Ma, Z.~Zheng, Z.~Luo, and Y.~Zheng, \enquote{Fabry-perot
  interferometric high-temperature sensing up to 1200 °c based on a silica
  glass photonic crystal fiber,} {{Sensors}} \textbf{18}
  (2018).

\bibitem{WangPengfei15}
P.~Wang, L.~Bo, Y.~Semenova, G.~Farrell, and G.~Brambilla, \enquote{Optical
  microfibre based photonic components and their applications in label-free
  biosensing,} {{Biosensors}} \textbf{5}, 471--499 (2015).

\bibitem{Lin12}
H.-Y. Lin, C.-H. Huang, G.-L. Cheng, N.-K. Chen, and H.-C. Chui,
  \enquote{Tapered optical fiber sensor based on localized surface plasmon
  resonance,} {{Opt. Express}} \textbf{20}, 21693--21701
  (2012).

\bibitem{Arissian14b}
L.~Arissian and J.-C. Diels, \enquote{Intracavity phase interferometry:
  frequency comb sensors inside a laser cavity,} {{Laser
  Photonics Rev}} \textbf{8}, 799--826 (2014).

\bibitem{Velten10}
A.~Velten, A.~Schmitt-Sody, and J.-C. Diels, \enquote{Precise intracavity phase
  measurement in an optical parametric oscillator with two pulses per cavity
  round-trip,} {{Optics Letters}} \textbf{35}, 1181--1183
  (2010).

\bibitem{Dennis91b}
M.~L. Dennis, J.-C. Diels, and M.~Lai, \enquote{The femtosecond ring dye laser:
  a potential new laser gyro,} {{Optics Letters}}
  \textbf{16}, 529--531 (1991).

\bibitem{Bohn97b}
M.~J. Bohn, J.-C. Diels, and R.~K. Jain, \enquote{Measuring intracavity phase
  changes using double pulses in a linear cavity,}
  {{Optics Lett.}} \textbf{22}, 642--644 (1997).

\bibitem{Lai92c}
M.~Lai, J.-C. Diels, and M.~Dennis, \enquote{Nonreciprocal measurements in fs
  ring lasers,} {{Optics Letters}} \textbf{17}, 1535--1537
  (1992).

\bibitem{Duling91}
I.~N. Duling, \enquote{All-fiber ring soliton laser mode locked with a
  nonlinear mirror,} {{Opt. Lett.}} \textbf{16}, 539--541
  (1991).

\bibitem{Nishizawa08}
N.~Nishizawa, Y.~Seno, K.~Sumimura, Y.~Sakakibara, E.~Itoga, H.~Kataura, and
  K.~Itoh, \enquote{All-polarization-maintaining Er-doped
ultrashort-pulse fiber laser using carbon
nanotube saturable absorber,} Optics Express \textbf{16}, 9429 (2008).

\bibitem{Set04}
S.~Y. Set, H.~Yaguchi, Y.~Tanaka, and M.~Jablonski, \enquote{Laser mode locking
  using a saturable absorber incorporating carbon nanotubes,}
  {{J. Lightwave Technol.}} \textbf{22}, 51 (2004).

\bibitem{Song07}
Y.-W. Song, S.~Yamashita, C.~S. Goh, and S.~Y. Set, \enquote{Carbon nanotube
  mode lockers with enhanced nonlinearity via evanescent field interaction in
  d-shaped fibers,} {{Opt. Lett.}} \textbf{32}, 148--150
  (2007).

\bibitem{Bao09}
Q.~Bao, H.~Zhang, Y.~Wang, Z.~Ni, Y.~Yan, Z.~X. Shen, K.~P. Loh, and D.~Y.
  Tang, \enquote{Atomic-layer graphene as a saturable absorber for ultrafast
  pulsed lasers,} {{Advanced Functional Materials}}
  \textbf{19}, 3077--3083 (2009).

\bibitem{Sotor15}
J.~Sotor, G.~Sobon, M.~Kowalczyk, W.~Macherzynski, P.~Paletko, and K.~M.
  Abramski, \enquote{Ultrafast thulium-doped fiber laser mode locked with black
  phosphorus,} {{Opt. Lett.}} \textbf{40}, 3885--3888
  (2015).

\bibitem{zhang09b}
M.~Zhang, L.~Chen, C.~Zhou, Y.~Cai, L.~Ren, and Z.~Zhang, \enquote{Mode-locked
  ytterbium-doped linear-cavity fiber laser operated at low repetition rate,}
  {{Laser Physics Letters}} \textbf{6}, 657--660 (2009).

\bibitem{guo14}
J.~Guo, \enquote{Bound-state solitons in a linear-cavity fiber laser
  mode-locked by single-walled carbon nanotubes,}
  {{Journal of Modern Optics}} \textbf{61}, 980--985
  (2014).

\bibitem{Bradley76}
I.~S. Ruddock and D.~J. Bradley, \enquote{Bandwidth-limited subpicosecond pulse
  generation in mode-locked cw dye lasers,} {{Appl. Phys.
  Lett.}} \textbf{29}, 296 (1976).

\bibitem{Fork81}
R.~L. Fork and C.~V. Shank, \enquote{Generation of optical pulses shorter than
  0.1 ps by colliding pulse mode-locking,} {{Appl. Phys.
  Lett.}} \textbf{38}, 671 (1981).

\bibitem{Zeng13}
C.~Zeng, X.~Liu, and L.~Yun, \enquote{{Bidirectional fiber soliton laser
  mode-locked by single-wall carbon nanotubes},} {{Optics
  Express}} \textbf{21}, 18937--18942 (2013).

\bibitem{Yao14}
X.~Yao, \enquote{Generation of bidirectional stretched pulses in a
  nanotube-mode-locked fiber laser,} {{Appl. Opt.}}
  \textbf{53}, 27--31 (2014).

\bibitem{Li18}
R.~Li, H.~Shi, H.~Tian, Y.~Li, B.~Liu, Y.~Song, and M.~Hu,
  \enquote{All-polarization-maintaining dual-wavelength mode-locked fiber laser
  based on sagnac loop filter,} {{Opt. Express}}
  \textbf{26}, 28302--28311 (2018).

\bibitem{Kieu08}
K.~Kieu and M.~Mansuripur, \enquote{All-fiber bidirectional passively
  mode-locked ring laser,} {{Opt. Lett.}} \textbf{33},
  64--66 (2008).

\bibitem{Krylov17}
A.~A. Krylov, D.~S. Chernykh, and E.~D. Obraztsova, \enquote{Colliding-pulse
  hybridly mode-locked erbium-doped all-fiber soliton gyrolaser,}
  {{Laser Physics}} \textbf{28}, 015103 (2017).

\bibitem{Masuda16}
K.~Masuda, J.~Hendrie, J.-C. Diels, and L.~Arissian, \enquote{Envelope, group
  and phase velocities in a nested frequency comb,}
  {{Journal of Physics B}} \textbf{49}, 085402 (2016).

\bibitem{Nicholson07b}
J.~W. Nicholson, R.~S. Windeler, and D.~J. DiGiovanni, \enquote{Optically
  driven deposition of single-walled carbon-nanotube saturable absorbers on
  optical fiber end-faces,} {{Opt. Express}} \textbf{15},
  9176--9183 (2007).

\bibitem{Afkhamiardakani16}
H.~Afkhamiardakani, B.~Kamer, J.~C. Diels, and L.~Arissian, \enquote{Carbon
  nanotubes for mode-locking: polarization study,} in \emph{Proceedings of
  Photonics West, Conference 9746-19,,} , vol. 9746 (SPIE, San Francisco,
  2016).

\bibitem{Afkhamiardakani18A}
H.~Afkhamiardakani, M.~Tehrani, and J.-C. Diels, \enquote{Extension of the
  stable operation of an all polarization maintaining mode-locked fiber laser,}
  in \emph{Conference on Lasers and Electro-Optics,}  (Optical Society of
  America, 2018), p. JTh2A.141.

\bibitem{Basov66}
N.~G. Basov, R.~V. Ambartsumyan, V.~S. Zuev, P.~G. Kryukov, and V.~S. Letokhov,
  \enquote{Nonlinear amplifications of light pulses,}
  {{Soviet Physics JETP}} \textbf{23}, 16--22 (1966).

\bibitem{Hendrie16}
J.~Hendrie, M.~Lenzner, H.~Afkhamiardakani, J.-C. Diels, and L.~Arissian,
  \enquote{Impact of resonant dispersion on the sensitivity of intracavity
  phase interferometry and laser gyros,} {{Optics
  Express}} \textbf{24}, 30402--304010 (2016).

\end{thebibliography}
\end{document}